\documentclass[prl,twocolumn,showpacs,preprintnumbers,amsmath,amssymb,floatfix]{revtex4}

\usepackage{graphicx}
\usepackage{bm}
\newcommand \be {\begin{equation}}
\newcommand \ee {\end{equation}}
\newcommand \ben {\begin{eqnarray}}
\newcommand \een {\end{eqnarray}}

\begin{document}

\title{
Melting at dislocations and grain boundaries: A Phase Field Crystal study
}

\author{Joel Berry$^1$, K. R. Elder$^2$, and Martin Grant$^1$}
\affiliation{$^1$ Physics Department, Rutherford Building, 3600 rue
University, McGill University, Montr\'eal, Qu\'ebec, Canada H3A 2T8}
\affiliation{$^2$Department of Physics, Oakland University,
Rochester, MI, 48309-4487} 


\date{\today}

\begin{abstract}
Dislocation and grain boundary melting are studied in three dimensions 
using the Phase Field Crystal method.
Isolated dislocations are found to melt radially outward from their core,
as the localized excess elastic energy drives
a power law divergence in the melt radius.
Dislocations within low-to-mid angle grain boundaries melt similarly until
an angle-dependent first order wetting transition occurs when neighboring
melted regions coalesce.
High angle boundaries are treated within a screening approximation,
and issues related to ensembles,
metastability, and grain size are discussed.
\end{abstract}


\maketitle

    Freezing and melting transitions do not exhibit the range of
universal behavior associated with continuous phase transitions and largely 
for this reason have eluded a unified theoretical description. The nature of
a given melting transition may depend sensitively on the details of the
system and experiment, and can involve
many distinct processes both within and between multiple forms of excitations.
For example, melting may occur abruptly and discontinuously at the 
melting temperature $T_m$, or it may initiate well below $T_m$ at surfaces
and/or internal defects and proceed up to $T_m$. This process
of premelting has been studied extensively for surfaces \cite{oxtoby90,lowen94}
and is relatively well understood, but limited and inconsistent experimental
evidence for melting at dislocations and grain 
boundaries leaves a number of issues unresolved.

    A recent study of colloidal crystals has verified that 
premelting does occur at vacancies, dislocations, and grain boundaries,
and has provided measurements
of the localized premelting behavior below $T_m$ \cite{yodhpremelt}. 
The conditions which determine whether premelting will occur continuously 
or discontinuously, and whether the
width of the premelted region diverges are not fully understood.
Grain boundaries in Al have been found to liquify very near $T_m$, and
the width of the melted layer appears to diverge \cite{hseih89}.
Discontinuous jumps in grain boundary diffusion coefficients \cite{budke99,
divinski05}, mobility \cite{author75}, and shear resistance \cite{watanabe84}
have been found in other metals.

    Theoretical studies have been based on either 
explicitly atomistic methods such as Molecular Dynamics 
\cite{yip92,broughton86} and Monte Carlo \cite{kikuchicahn80}, 
or on continuum phase field models with uniform phases 
\cite{tang06,warren02,rappaz02}.
In this study, dislocation and grain boundary melting are examined using
the Phase Field Crystal (PFC) method \cite{pfc04}, which extends the
phase field approach to the level of atomistic resolution.
This permits straightforward identification of stable equilibrium and metastable
non-equilibrium atomic structures, while inherently including
crystal symmetry and orientation, elasticity/plasticity, and the 
individual dislocations which compose the grain boundaries. 
Our description will be most applicable to hard-sphere/colloidal systems
and possibly simple metals.

    The melting behavior of dislocation pairs and
symmetric tilt grain boundaries of $\theta=4^{\circ}$,
$8^{\circ}$, $16^{\circ}$, $24^{\circ}$, $32^{\circ}$, and $44^{\circ}$
are examined numerically for a simple PFC model with bcc symmetry.
Analytic results are derived for isolated dislocations and low
$\theta$ boundaries by combining the PFC equations with continuum linear 
elasticity.
A screening approximation is outlined for high angle boundaries, though 
somewhat surprisingly, the low $\theta$ description is found to remain
reasonably accurate for high $\theta$.

    As shown in reference \cite{pfcdft07} the PFC free energy can be derived 
from the Ramakrishnan-Yussouff free energy functional of classical density 
functional theory \cite{ry79,s91}. Here we give the final form:
\be
F=\int d\vec{r}\bigg[
\frac{B^\ell}{2} n^2+\frac{B^s}{2} n(2\nabla^2+\nabla^4)n 
- v\frac{n^3}{6}+\frac{n^4}{12} 
\bigg]  
\label{eq:fpfc}
\ee
where $F\equiv ({\cal F}-{\cal F}_o)/(\bar{\rho}k_BT L^d)$, 
${\cal F}_o$ is the free energy functional at constant density,
$\bar{\rho}$ is the average number density, 
$k_B$ is Boltzmann's constant, $T$ is temperature, 
$L\equiv \sqrt{2|\hat{C}_4|/\hat{C}_2}$,
$C_i$ is the $i$ point direct correlation function of the reference
liquid state,
$B^{\ell} \equiv 1-\bar{\rho} \hat{C}_0$, 
$B^{s} \equiv \bar{\rho} (\hat{C}_2)^2/4|\hat{C}_4|$, 
$\vec{r} = \vec{x}/L$ and 
$v$ accounts for the lowest order contribution 
from $C_3$. 
$n=(\rho-\rho_L)/\rho_L$ is the scaled number density, where
$\rho$ is the local density variable and $\rho_L=\bar{\rho}$ is the liquid
number density.
Classical density functional theory has been used to examine surface melting 
\cite{ohnesorge94,lowen94}, but not grain boundary melting,
presumably due to the complexity of the
solid-solid interface and the more demanding system size requirements.

    The dynamics are given in dimensionless form by
\be
\frac{\partial n}{\partial \tau}
 = \nabla^2 \frac{\delta F}{\delta n} + \eta
\label{eq:dpfc}
\ee
where $\langle\eta(\vec{r}_1,\tau_1)\eta(\vec{r}_2,\tau_2)\rangle =
M \nabla^2\delta(\vec{r}_1-\vec{r}_2) \delta(\tau_1-\tau_2)$ 
and $M \equiv 3(\bar{\rho}L^d)^2$. This form imposes a constant density 
and is consistent with the canonical ensemble.

    A semi-implicit pseudospectral algorithm was used to solve Eq.
(\ref{eq:dpfc}) in three dimensions with periodic boundary conditions.
The parameters used were $\Delta x=0.976031$, $\Delta \tau=0.5$,
$B^s=\sqrt{3}/3$, $v=3^{1/4}/2$, and $M=0.002$. 
A system size $V=(512\Delta x)^3=(56a)^3$ was used for the
dislocation pair and $4^{\circ}$ grain boundary pair, while
$V=(256\Delta x)^3=(28a)^3$ was used for all other grain boundary pairs,
where $a=8.9237$ 
is the bcc lattice constant.
Finite size effects increase as $\theta$ decreases, but were found to
be small for all grain boundaries studied.
The gaussian width or mean square displacement (D) of each localized 
density peak was monitored as the temperature
$\Delta B \equiv B^{\ell} - B^s$
was increased toward the melting point. The local crystallinity $\phi$ 
has been defined as
$\phi(\vec{r})=(D(\vec{r})-D_X)/(D_L-D_X)$, 
where $D_X$ is the equilibrium D of the crystal phase and $D_L$
is that of the liquid phase. 
$\phi=1/2$ specifies a liquid-solid interface.

\begin{figure}
\includegraphics[width=80mm]{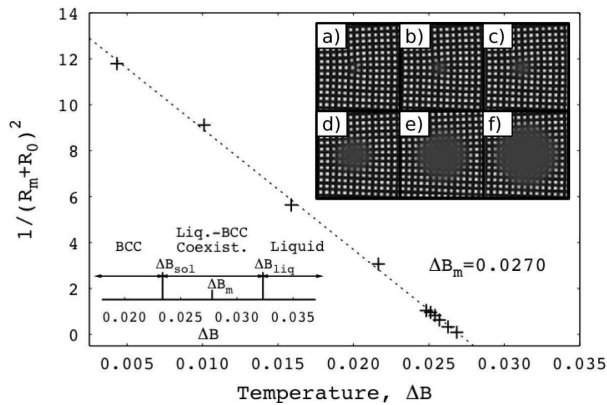}
\caption{Numerically measured local melt radius $R_m$ 
around an edge dislocation in a bcc crystal as a function of temperature
(units of lattice constant $a$, $\Delta B < 0.025$ values obtained by 
extrapolation of $\phi(\vec{r})$). 
$R_0=0.2812a$ is the zero temperature radius, determined by best fit.
Inset: cross-sectional images of $n(x,y,z)$ from simulations at 
$\Delta B=0.02165,0.02511,0.02569,0.02627,0.02656,0.02685$ from a) to f),
showing melting at a dislocation core.
}
\label{Rmelt}
\end{figure}

    The radius of melted region around a dislocation core $R_m$ was first
measured in this manner for an edge dislocation pair as the temperature 
was raised toward the bulk melting temperature $\Delta B_m$.
The results are shown in Fig. \ref{Rmelt}, where
the data is plotted as $(R_m+R_0)^{-2}$ vs $\Delta B$ to demonstrate
that $R_m$ is consistent with a $(\Delta B_m-\Delta B)^{-1/2}$ form which 
will be derived later. $R_0$ is an offset related to the finite
size of the dislocation core at zero temperature. The fit to this form 
predicts a bulk melting temperature $\Delta B_m=0.0270$ which is in good
agreement with the directly measured value of $\Delta B_m=0.0278$.
The upper inset of Fig. \ref{Rmelt} shows melting around an edge
dislocation as $\Delta B \rightarrow \Delta B_m$.

\begin{figure}
\includegraphics[width=80mm]{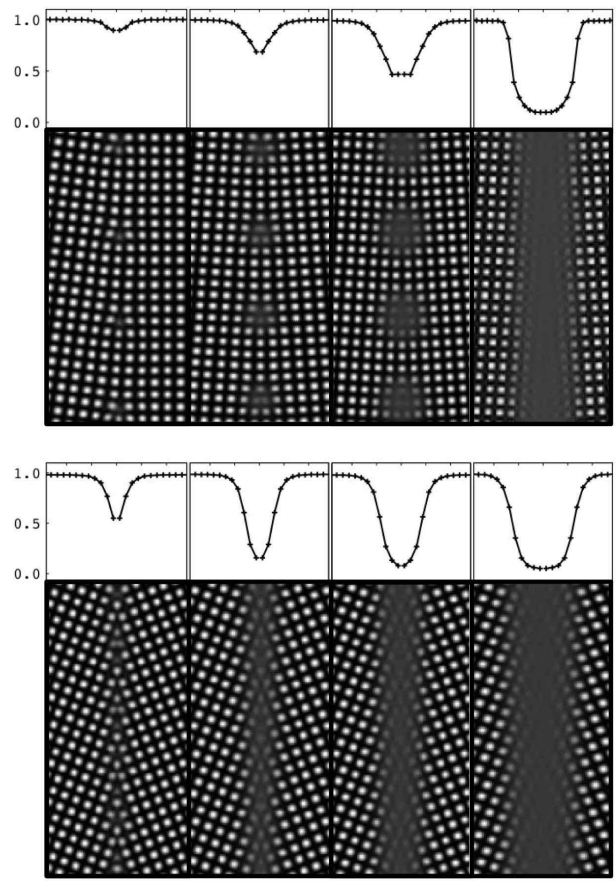}
\caption{Laterally averaged crystallinity parameter $\phi$ and cross-sections
of $n(x,y,z)$ for $8^{\circ}$ (upper) and $44^{\circ}$ (lower) grain 
boundaries at $\Delta B=0.02165,0.02627,0.02656,0.02685$ and 
$\Delta B=0.02425,0.02468,0.02540,0.02656$ from left to right, respectively.
}
\label{gb2}
\end{figure}

    Figure \ref{gb2} shows the progression of melting at $8^{\circ}$ and
$44^{\circ}$ grain boundaries. Low angle boundaries were found to first melt
radially at each dislocation core, until the melted regions of neighboring 
dislocations coalesce and a uniform wetting layer is formed along the boundary.
Individual dislocations cannot be distinguished in high angle boundaries,
and melting in this case was found to proceed by uniform disordering along 
the boundary rather than by local radial melting. Interfacial roughening
due to thermal fluctuations is negligible in all simulations presented here. 

\begin{figure}
\includegraphics[width=80mm]{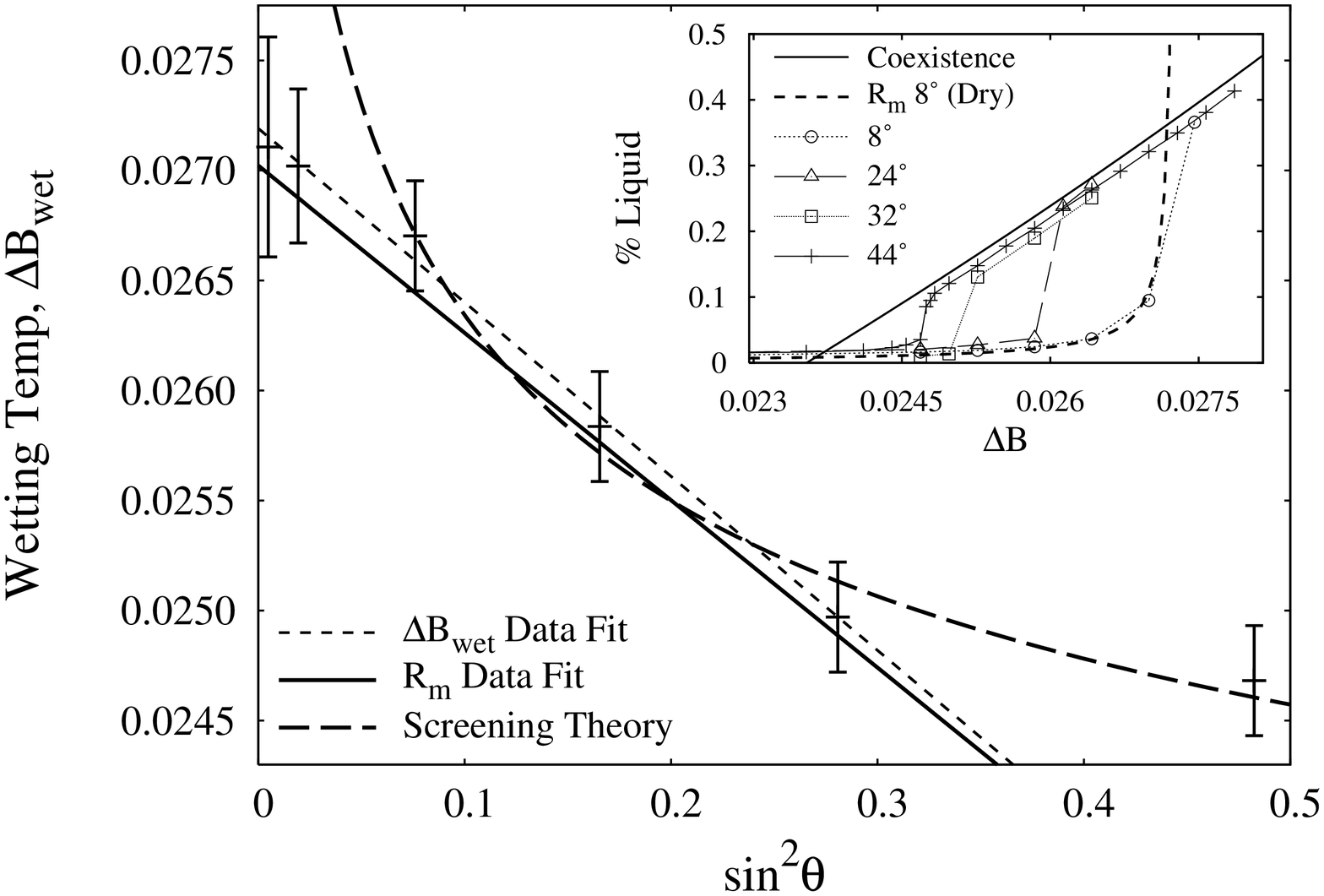}
\caption{Grain boundary wetting temperature vs. $\sin^2{\theta}$. 
Fit lines are discussed in the text.
Inset: Liquid volume fraction vs $\Delta B$ for a various grain boundaries.
The dotted line corresponds to the liquid volume fraction predicted for
the $8^{\circ}$ boundary based on Eq. (\ref{radialr}) only (ignoring
coexistence).}
\label{Twet}
\end{figure}

    The dependence of the width of the wetting layer (or the liquid volume
fraction of the system) on $\Delta B$ is shown in the inset of Fig.
\ref{Twet} for various grain boundary angles. In all cases the width
remains narrow and the boundary relatively dry until above the solidus, 
at which point a discontinuous
jump is observed at some characteristic wetting temperature $\Delta B_{wet}$.
The dependence of $\Delta B_{wet}$ on $\theta$ is shown in the main plot
of Fig. \ref{Twet}. The fit lines will be discussed in the following, 
though the axes of the plot reveal already that our predicted form will be
$\Delta B_{wet} \propto \sin^2{\theta}$.

    Based on these simulation results, we have developed a theory
of dislocation-driven melting, which is easily extended to low angle grain
boundaries. The low angle results are shown to remain accurate
for all but the highest $\theta$ where the dislocation spacing $d$ reaches
the order of the burger's vector $b$. A screening approximation for the
spatial grain boundary energy is found to
be more applicable for very large $\theta$, with a gradual crossover
taking place between these two regimes.

    According to continuum elasticity \cite{chaikluben}, the radially averaged
elastic energy
density per length of dislocation line in a three-dimensional isotropic solid
is $\bar{F}_{el}=\alpha \mu/R^2$, where $\mu$ is the shear modulus.
For a screw dislocation 
$\alpha_s=b^2/4\pi^2$ and for an edge dislocation 
$\alpha_e=\alpha_s/(1-\sigma)$ where $\sigma$ is the Poisson ratio.
If we assume this result to hold for a dislocation in the PFC model, at
distances approaching the core region, then $R_m$ can be directly calculated
by determining the distance at which $\bar{F}_{el}$ is sufficiently large to
destabilize the crystalline phase, melting the dislocation core.

    It will be assumed that $n$ can be represented in a one mode approximation 
for a bcc lattice, i.e., 
$n(\vec{r})=A\left(\cos{qx}\cos{qy}+\cos{qx}\cos{qz}+\cos{qy}\cos{qz}\right)$.
Substituting $n$ into Eq. (\ref{eq:fpfc}) and minimizing with 
respect to $q$ gives, 
\ben
\Delta f^X &=& \frac{3}{8}\Delta B A^2 -\frac{v}{8}A^3 + \frac{45}{256}A^4  
\een
where $\Delta f^X\equiv (F-F_L)/V$, $F_L$ is the free energy 
of the liquid, 
$V=(2\pi/q)^3$, and $q_{min} = \sqrt{2}/2$.  
The shear modulus can be estimated in 
the one mode approximation by setting $n(x,y,z)\rightarrow n(x+\zeta y,y,z)$ 
and expanding $F$ in $\zeta$ such that $F = F(\zeta=0) + \mu \zeta^2+ 
\cdots$.  This procedure gives $\mu/k_B T L^2\bar{\rho} = A^2 B^s/8$.  
The total (dimensionless) free energy of the system with a dislocation is 
then $\Delta f^X + \bar{F}_{el}/k_B T \bar{\rho} L^d$, which can be written 
\ben
\Delta f = \frac{3}{8}\left(\Delta B+\frac{E}{\bar{R}^2}\right) A^2 
-\frac{v}{8}A^3 + \frac{45}{256} A^4.
\label{eq:fen}
\een
where $\bar{R}\equiv R/b$ and $E_s\equiv B^s/(12\pi^2)$, 
$E_e\equiv E_s/(1-\sigma)$ for screw 
and edge dislocations respectively. Equation 
(\ref{eq:fen}) indicates that the elastic energy `shifts' the effective 
temperature $\Delta B$ by an amount $E/\bar{R}^2$.  The implication 
is that the liquid-solid transition is shifted and instead of occurring 
when $\Delta B = \Delta B_m$ occurs when 
when $\Delta B+E/\bar{R}^2 = \Delta B_m$.  Thus 
the premelt radius can be written
\be
\bar{R}_m = \sqrt{E/(\Delta B_m-\Delta B)}
\label{radialr}
\ee
or $1/\bar{R}_m^2 = (\Delta B_m-\Delta B)/E$.

    As shown in Fig. \ref{Rmelt}, this form
is consistent with the simulation results for edge dislocation pairs, though
the predicted slope ($-1/E_e$) is in error by a factor of nearly five. A more 
definitive test would require additional data very near $\Delta B_m$,
a region increasingly difficult to access due to system size requirements.

    Extending this result to low angle boundaries, we can estimate
the grain boundary wetting temperature $\Delta B_{wet}$ where neighboring
dislocations coalesce by setting $\bar{R}_m=d/2=1/(2\sin{\theta})$.
Substituting Eq. (\ref{radialr}) for $\bar{R}_m$ gives 
\be
\Delta B_{wet}=\Delta B_m - 4E\sin^2{\theta}
\label{Tweteq}
\ee
which is in good agreement with the data
shown in Fig. \ref{Twet}. This result should lose accuracy as $\theta$
increases since the dislocation energies gradually deviate from the isolated
dislocation result as $\theta \rightarrow \theta_{max}$. The agreement
up to $\theta \simeq 32^{\circ}$ is somewhat unexpected as the superposition
generally loses accuracy for $\theta \gtrsim 10^{\circ}$. The best fit
line predicts $\Delta B_m=0.0272$, again near the measured value.

    The solid line in Fig. \ref{Twet} corresponds to the fit line from 
Fig. \ref{Rmelt}, set equal to $1/(2\sin{\theta})$ and solved for
$\Delta B_{wet}$. The agreement here clearly indicates that the wetting
of low and mid-angle grain boundaries is accurately described by the
coalesence of radially melted regions around nearly isolated dislocations.
    
    In the limit of large $\theta$ ($d \rightarrow 0$), the grain boundary
energy becomes increasingly uniform along its length (see Fig. \ref{gb2}) 
and can no longer be described in terms of individual dislocations.
We expect that elastic fields at long distances are screened by closely
spaced dislocations, giving rise to exponentially decaying spatial grain
boundary energy.  Indeed, direct analysis of free energy data from
simulations indicates that such an exponential form is qualitatively correct.
Solving for $\Delta B_{wet}$ using
$\bar{F}_{el} \propto {\rm e}^{-x/b}$ rather than $\bar{F}_{el} \propto 1/R^2$
gives $\Delta B_m - \Delta B_{wet} \propto {\rm e}^{-f(\theta)}$. This
is the form of the wide dashed line in Fig. \ref{Twet}, which appears to be
more appropriate for large $\theta$. 

    Some comments concerning the influence of liquid-solid coexistence and the 
canonical ensemble (i.e., conserved density) on grain boundary melting may 
be helpful at this point.
The equilibrium state for a simple system with a grain boundary is
most generally either dry if $F_{gb} < 2F_{ls} +
\ell(\bar{F}_L - \bar{F}_X)$ or wet if 
$F_{gb} > 2F_{ls} + \ell(\bar{F}_L - \bar{F}_X)$, where 
$F_{gb}$ is the grain boundary energy, $F_{ls}$ is
the energy of a liquid-solid interface, and $\ell$ is the width
of the liquid region in the wet state. If the wet state becomes favorable
below the melting temperature, then a grain boundary wetting transition occurs.
In the canonical ensemble as examined here, the effects of 
liquid-solid coexistence and the subsequent shifts in density of the two
phases above the solidus $\Delta B_{sol}$ modify this heuristic argument. 
Now $\Delta B_m$, the
temperature at which $\bar{F}_L=\bar{F}_X$, is straddled by a
coexistence region. As $\Delta B \rightarrow \Delta B_m$
the system first encounters a solidus 
above which some volume fraction of
liquid will minimize the overall $\bar{F}$. For the grain boundary pair
geometry, the equilibrium state above $\Delta B_{sol}$ is one with a uniform
volume of liquid occupying each boundary region. 
Therefore, an equilibrium first order wetting transition will occur
at $\Delta B_{sol}$ as long as the grain size is not excessively small.
Above $\Delta B_{sol}$, the liquid layer width will be 
controlled by coexistence rather than local defect energies,
since the elastic fields of the grains largely decouple 
($\bar{F}_{el} \rightarrow 0$) upon wetting.

    The results presented here show no wetting or strong premelting 
for $\Delta B \le \Delta B_{sol}$, and the equilibrium wetting transition 
is not observed. Instead, a $\theta$-dependent discontinuous transition from 
the metastable dry boundary state to the equilibrium wet state occurs at 
$\Delta B_{sol} < \Delta B_{wet} < \Delta B_m$, as shown
in the inset of Fig. \ref{Twet}.
This is because the wetted state is not nucleated in 
observable times until $R_m$ has grown sufficiently large to coalesce
and the free energy barrier approaches zero.
The dislocations and/or grain boundaries act as
nucleation sites for the liquid above $\Delta B_{sol}$, creating
well-defined nonequilibrium paths from the metastable dry 
state to the $F$ minimizing wet state
(which all must conserve $\rho$). 

    The condition for wetting in the canonical ensemble involves the grain
size $L_g$, such that wetting can be supressed to temperatures above 
$\Delta B_{sol}$ when $L_g$ is finite. The condition can be written
$F_{gb} + L_g\Delta F_X > 2F_{ls} + \ell \Delta F_C$ where
$\Delta F_X = \bar{F}_X[\bar{\rho}] - \bar{F}_X[\rho_X]$ and
$\Delta F_C = \bar{F}_L[\rho_L] - \bar{F}_X[\rho_X]$.
Here $\ell=(\rho_X-\bar{\rho})/(\rho_X-\rho_L)$, 
$\bar{\rho}$ is the conserved system density,
and $\rho_X$ and $\rho_L$ are the shifted coexistence densities of the solid
and liquid phases, respectively.
For $\Delta B \le \Delta B_{sol}$, if we assume that $\Delta F_X=0$ and 
$\rho_L=\rho_X=\bar{\rho}$
(this is not the case when premelting is strong below $\Delta B_{sol}$),
we recover the original inequality $F_{gb} > 2F_{ls} +
\ell(\bar{F}_L - \bar{F}_X)$ and
$L_g$ is not a significant factor. 
In the limit $L_g \rightarrow \infty$, the wetting condition will always be
satisfied for $\Delta B > \Delta B_{sol}$ and the equilibrium
transition occurs at $\Delta B_{sol}$. As $L_g$ decreases, the equilibrium
wetting transition is shifted to higher $\Delta B$.

    Two-dimensional systems show qualitatively similar results to those
presented above.
Preliminary simulations have also been conducted in the grand canonical
ensemble (i.e., non-conserved density), 
where the above complications with liquid-solid
coexistence due to density conservation can be avoided. Initial results
indicate that there is no strong wetting transtion in this case, as the
atoms near grain boundaries continuously delocalize toward $\Delta B_m$
but do not liquify and decouple the grains until above $\Delta B_m$.

    JB would like to acknowledge support from the Richard H. Tomlinson 
and Carl Reinhardt Foundations of McGill University and thank
Maria Kilfoil, Dan Vernon, Zhi-Feng Huang,
Katsuyo Thornton, and Nilima Nigam for useful discussions.
KRE acknowledges support from the
National Science Foundation under Grant No. DMR-0413062.
MG acknowledges support from the Natural Sciences and Engineering
Research Council of Canada, and {\it le Fonds qu\'{e}b\'{e}cois  de la
recherche sur la nature et les technologies\/}.

\end{document}